\documentclass[journal]{IEEEtran}

\ifCLASSINFOpdf
   \usepackage[pdftex]{graphicx}
\else
\fi
\begin{document}
%
\title{Hybrid architecture for satellite data processing workflow management}
%
%
%

\author{M. Naresh Kumar,

\thanks{M. Naresh Kumar at National Remote Snsing Centre, Hyderabad,
Telangana, 500 037 India e-mail: (nareshkumar\_m@nrsc.gov.in).}
}

\maketitle

\begin{abstract}
The ever growing demand for remote sensing data products by user community has resulted in many Indian and foreign remote sensing satellites being launched. The diversity in the remote sensing sensors has resulted in heterogeneous software and hardware environments for generating geospatial data products. The workflow automation software knows as information management system is in place at National Remote Sensing Centre (NRSC) catering to the needs of the data processing and data dissemination. The software components of workflow are interfaced in different heterogeneous environments that get executed at data processing software in automated and semi automated modes. For every new satellite being launched, the software is modified or upgraded if new business processes are introduced. In this study, we propose a software architecture that gives more flexible automation with very less manageable code.  The study also addresses utilization and extraction of useful information from historic production and customer details. A comparison of the current workflow software architecture with existing practices in industry like Service Oriented Architecture (SOA), Extensible Markup Languages (XML), and Event based architectures has been made. A new hybrid approach based on the industry practices is proposed to improve the existing workflow.
\end{abstract}


%
\IEEEpeerreviewmaketitle

\section{Introduction}
\begin{figure}[!t]
\centering
\includegraphics[width=3in]{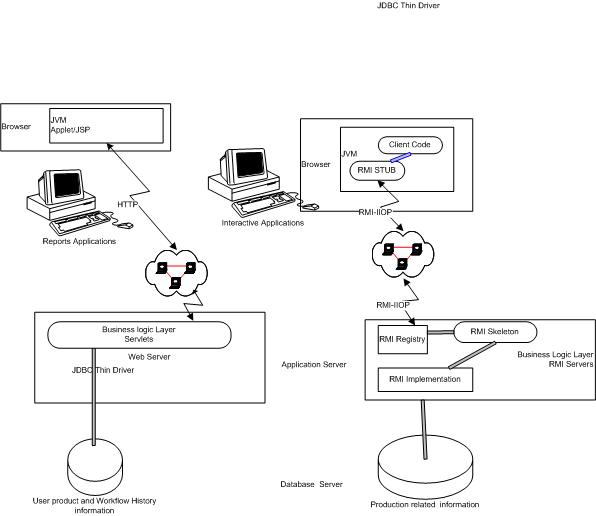}
 \caption{Generalized Architecture of Workflow automation software at NRSA}
\label{fig1}
\end{figure}

%
%
%
%
\IEEEPARstart{D}{ata} processing area of National Remote
Sensing Centre (NRSC) is responsible for generation of remote sensing data products for Indian (IRS) and foreign remote sensing satellites. Data product generation involves interfacing different business processes at User request processing, Data processing, Value addition, Film generation, Photo processing and Quality control. These processes have been converted to information technology processes by the information management system (IMS) software. The software identifies every process as a work center or group of work centers.  The product generation is based on the type of satellite and processing level that is used to route the products to different work centers based on the concept of work order or job order.   The contents of work order are available virtually as and when required to different nodes of work centers to process in an automated mode or through human interaction. It is noted that during the life time of a product generation, the contents of the work order change, i.e., the contents are dynamic. The IMS application is based on three tier software architecture with database layer under Oracle 8i environment, business layer and presentation layers that are implemented using J2EE technologies as shown in Figure \ref{fig1} providing the flexibility to replace any layer with out affecting the other layers \cite{5}. The IMS design has been accomplished using object oriented methodologies (which provides the mechanism of capturing the functionality and the data associated with real object \cite{12,14} and is implemented in Java.  Exchange of data between presentation layer and business layer is through business objects. The data is formatted for presentation before displaying it on to the browser, wherever human intervention or monitoring is involved.

\begin{figure}[!t]
\centering
\includegraphics[width=3in]{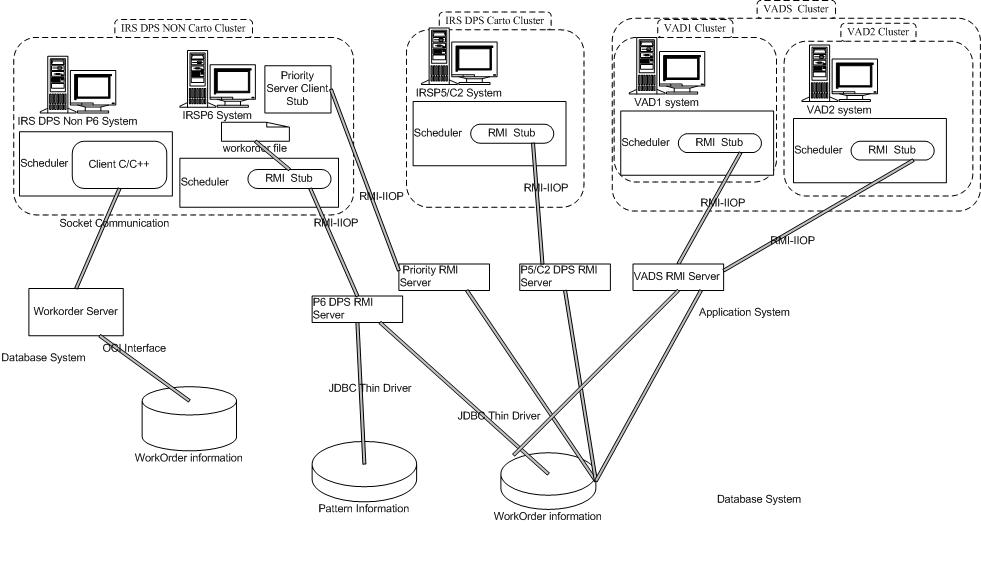}
 \caption{Data exchange mechanisms between the IMS software and interactive, automated work centers}
\label{fig2}
\end{figure}

The data is also provided to automated processes like data processing schedulers that generate data products by ingesting user inputs that are a part of the work order, and ancillary data of the satellite as shown (ADIF) in Figure \ref{fig2}.

A transfer of data to presentation layer and data processing system clients is accomplished by data marshaling and un-marshaling using java remote method invocation \cite{13}. These interactions are tightly coupled resulting in high maintenance of code for every change in a business process. The work order flow from system to system is based either on interactions through a browser or requests from and automated process i.e., the scheduler. The transfer of data is highly synchronous and happens as blocking and unblocking calls in the client and server processes resulting in slowing down of the system during heavy data transfers. The IMS software stores online transactions between the work-centers in an operational environment and historic transactions in a lightly summarized manner. Being a centralized repository of data, management looks for information for decision making from database managed by IMS software. Therefore building a dataware house environment would help to build and realize the decision making tools.

The paper is divided in to the following sections. In Section 2 we highlight the adaptation of service oriented architecture including event based process communication for IMS is highlighted. In Section 3 XML based data exchanges between the different layers are discussed. In Section 4 data warehouse schemes are discussed. In Section 5 hybrid architecture for workflow management is proposed and Section 6 conclusions are presented that summarizes the benefits of proposed hybrid architecture for workflow automation of data products generation.

\section{Service Orientation and Event based communication}

The aim of service oriented Framework is to provide a loose coupling between operating systems, programming languages and other technologies which underlie applications \cite{11}. SOA separates functions into distinct units, or services \cite{1}, which are made accessible over a network in order that they can be combined and reused in the production of business applications \cite{6}. SOA is an architecture that relies on service-orientation as its fundamental design principle. In a SOA environment, independent services can be accessed without knowledge of their underlying platform implementation \cite{4}. SOA relies on services exposing their functionality via interfaces which other applications and services read to understand how the service can be utilized.

Workflow automation requires functionality like work-order generation, updating and dispatch which are similar for different work centers. The work-centers interface with these software interfaces for automating the tasks in the production chain. Also the work-centers involved may be generating data products heterogeneous systems. To continue the usage of the existing and new systems in the production these software interfaces can be coined as services and service orientation will permit the usage of different production hardware and software environment to coexist. Also the development and maintenance of the software modules is reduced by adopting the service oriented architecture.

The communication between the software modules across the data products generation systems is done using object oriented messaging. This has resulted in a tightly coupled software communication interfaces. The present industry standard to reduce the coupling between the software modules is the Event-driven architecture (EDA). The EDA is a software architecture pattern promoting the production, detection, consumption of, and reaction to events. The architectural pattern is applied for design of applications and systems which communicate using loosely coupled software components and services. An event-driven system typically consists of event emitters (or agents) and event consumers (or sinks). Sinks have the responsibility of applying a reaction as soon as an event is presented. The reaction might or might not be completely provided by the sink itself. For instance, the sink might just have the responsibility to filter, transform and forward the event to another component or it might provide a self contained reaction to such event. The first category of sinks can be based upon traditional components such as message oriented middleware while the second category of sinks (self contained online reaction) might require a more appropriate transactional executive framework. Event-driven architecture can complement service-oriented architecture (SOA) because services can be activated by triggers fired on incoming events \cite{3,8}. This paradigm is particularly useful whenever the sink does not provide any self-contained executive.

\begin{figure}[!t]
\centering
\includegraphics[width=3in]{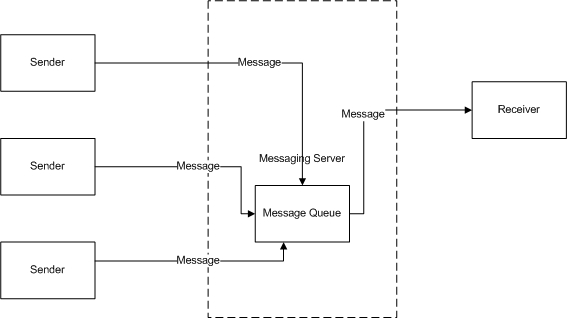}
 \caption{Messaging passing between the sender and receiver using message queue on the messaging server}
\label{fig3}
\end{figure}

There are two types of control namely centralized and event based which determines the ordering and rules for execution of the component. There are three types of coupling in the centralized control referential, communication and execution. The referential coupling requires the sender to know the receiver in advance before communication. The communication coupling requires the sender and receiver to activate at the same time. The execution coupling the sender must wait for the receiver to complete the message processing. The event based communication provides the features like dependency inversion, concurrency and message queuing. The dependency inversion reverses the referential coupling, concurrency removes the execution coupling and message queuing removes the communication coupling. The temporal decoupling between the sender and the receiver can be achieved by implementing a messaging server Figure \ref{fig3}.

\section{XML based data exchange mechanism}
The present design of data exchange is based on sending serialized objects and ASCII file based mechanisms to different automated and non automated work-centers. The serialization and de-serialization in the remote procedure calls (RPC) results in increased communication between the client and server systems. Also, the change in the object would results in a different version of the software. The ASCII based mechanism provides cross platform support but the volume of data to be transferred would increase and leads to communication overheads. The present industry standard for data exchange is extensible markup language (XML).

XML is a user-driven, open standard for exchanging data both over corporate networks and between different enterprises, notably over the Internet. XML's biggest potential lies in its ability to mark up mission-critical document elements self-descriptively. XML transports the metadata (the information about the data) together with the relevant data, thus allowing its meaning to be easily interpreted. In addition, XML enables suitably coded documents to be read and understood without difficulty by both humans and machines. XML as a data interchange format is compelling, primarily because it gives developers, a language with which to more easily identify interoperability problems and a common syntax and tool set with which to fix them.
The XML related standards like eXtensible Stylesheet Language (XSL), XSL transformations (XSLT) and document type definition (DTD), provide the mechanisms for styling the web pages, validating the data and verifying the well-formed ness of the data respectively. The styling using XSL is external to data and is independent of any software as is being currently done using Servlets/JSP generating HTML tags. Different look and feel can be provided with out any additional software just by configuring the XSL files for report generation.

The data from databases can be easily retrieved from the databases like Oracle using XML SQL Utility (XSU). The XSU uses a schematic mapping that defines how to map tables and views, including object-relational features, to XML documents. Oracle translates the chain of object references from the database into the hierarchical structure of XML elements. The XSU can also be used for executing queries in a Java environment and retrieve XML from the database. XML file exchange mechanism is platform independent and also has different mechanism for styling, validating and other features making it attractive approach for data exchanges across platforms and loose coupling of software modules.

The present file based using client/server based software involves lot of development and maintenance cost. Also, the presentation (look and feel) of the data for report generation requires addition software to be developed. Adopting XML based exchange mechanism would reduce the maintenance costs of the software modules.

\section{Data Warehouse Environment}

IMS software uses Oracle 8i database which is an operational database optimized for transaction management. NRSA data users and other work-center users involved in the data products generation constantly require historic information in the form of reports like turn around time (TAT) for each work center, products wise TAT , completed products between any two dates, pending analysis reports etc.,.  The present database being fine tuned for transaction management and normalized to a high degree resulting in overheads on the database server. Analysis on the production data can be better exploited using a data warehouse environment. A Data warehouse environment provides drill down and roll up kind of analysis which is the crux for applications like customer relation management.

A data warehouse is the repository of the entire organizational data and is designed to optimize reporting and analysis of the archived data \cite{7}. The operational databases are optimized for data integrity and speed of recording the business transactions with the use of relational models and normalization.  The relational databases are efficient in managing the relationships between the tables and they are fast in inserts and updates when dealing with less volume of data.  However, they become slower as the data size increase.

In a data warehouse, data is gathered from operational systems and retained even after the same has been purged from operational systems.  The operational data is lightly summarized and modeled as facts and dimension table. Data is often stored multiple times - in their most granular form and their summarized forms are called aggregates.  Further, the data is de-normalized by dimensional data modeling \cite{9} that in turn facilitates a very quick retrieval of the same.

The data warehouse can be designed in two ways knows as star schema and snowflake schema. The snowflake schema the dimension tables are further normalized which is not the case in star schema. The snow flake schema reflect the way in which users think about data therefore the drill down and roll up analysis can be easily performed \cite{10}. The snowflake models are intuitive and easy to understand, amenable to query optimization since arbitrary n-way joins with the fact table can be evaluated by a single pass through the fact table, can accommodate for aggregate data, and are easily extensible by adding new attributes to the fact table or to one or more of the dimension tables and new dimension tables to the schema without interfering with existing database programs.
\begin{figure}[!t]
\centering
\includegraphics[width=3in]{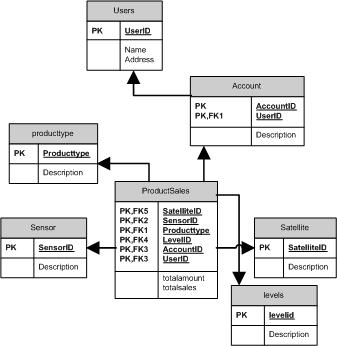}
 \caption{A snowflake schema model for product sales}
\label{fig4}
\end{figure}

An example of data ware modeling would be a snowflake schema for remote sensing data product sales. The remote sensing data user product sales can be modeled as a fact and dimensions could be customer, satellite, sensor, product type, level of correction. The schema shown in the Figure \ref{fig4}  optimizes the queries for finding the summary of total amount of sales in different dimensions. If the management wants to know the total sales and total amount for each dimension can be easily retrieved from the data ware house because the total amount and sales are already pre computed for different dimensions. The star schema is easily convertible to snow flake schema by normalizing all the dimension tables. The Oracle database supports the snowflake schema which would improve the time required for data analysis. The snow flake schema design is user specific, i.e. data analysis is optimized from the perspective of the management.

The data warehouse building design is to be undertaken in planned and phased manner. A corporate data model is developed keeping in view the requirements of an analyst. The extract, transform and load programs (ETL) are to be executed as when the wrinkle time (the time data will not be used further updated in the database) of the data is reached in the operational database which is normally 24 hours for standard products. These are some of the overheads for building the data ware house environment. For ever changing business orientation the analysis can be easily done using the data ware house environment than the operational databases.

\section{Proposed Hybrid Architecture}
The data products generation at NRSC involves many heterogeneous system with operating system platform generating data products of all the remote sensing satellite right from IRS 1A to Cartosat-2. The data products for IRS1A to IRS-P6 satellite are being generated on SGI systems on IRIX operating system. The IRS P6 AWIFS products are generated on PC based Linux OS platform. The IRS P5 and Cartosat-2 products are being generated on PC based Linux platforms. In addition, there are two value additions work-centers supporting more than ten systems for value addition to the remote sensing data products. The IMS software interfaces provide services like ADIF and work-order to all data products generation and value addition systems.  The other work-center users' interact with IMS software for product entry and monitoring using reports using Browser based services. The IMS software currently handles more than 100 products per day in the production chain. Considering the heterogeneity of the platforms, operating systems, complexity of the application domains and interfaces to associated processes, a hybrid approach to software architecture simplifies realization of interoperable machine independent interfaces and provides extensibility for future requirements.

Considering the advantages of technologies discussed in above sections, we propose a hybrid software architecture  that is services oriented, facilitating data exchanges using XML and message exchanges between the modules using event mechanisms, as refinements to the existing IMS architecture. These approaches would enable loose coupling between the software modules. The software maintenance and deployment costs are minimized because of the reduced dependencies between the software modules and platforms.  

\begin{figure}[!t]
\centering
\includegraphics[width=3in]{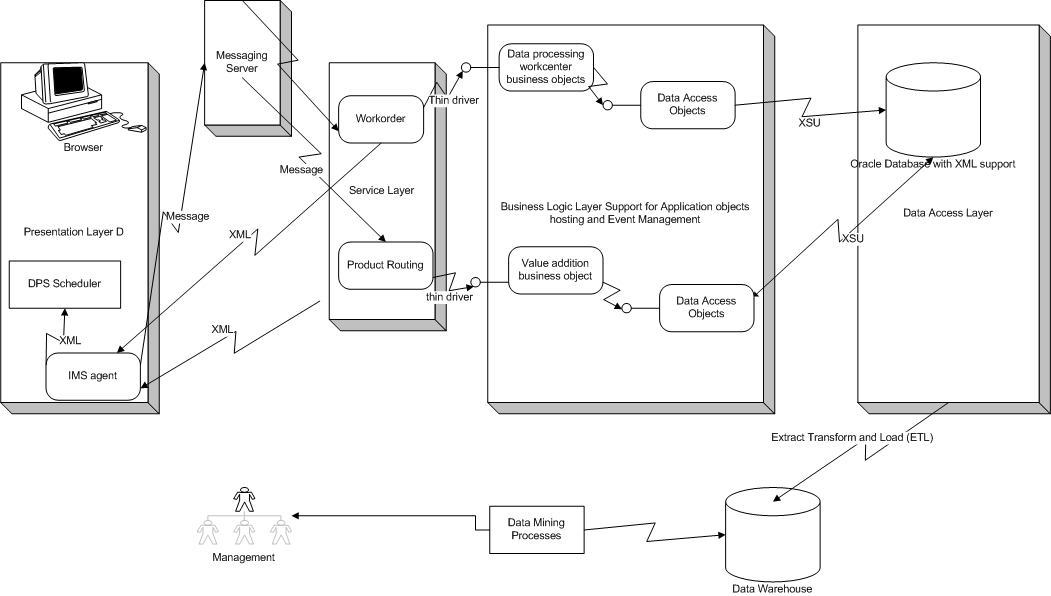}
 \caption{Proposed hybrid architecture for work flow automation software for monitoring data product generation}
\label{fig5}
\end{figure}

The hybrid architecture (Figure \ref{fig5}) also supports a simultaneous operation of existing computationally intensive data products generation systems with an option to reengineer the processes with state of art plug and play technologies.

A new layer known as service layer is introduced between the business layer and the presentation layer. The service layer provides necessary abstraction of underlying mechanisms for providing workflow operations such as work-orders, product routing etc., A data exchange mechanism using XML removes nativity of the data associated with the system i.e. it becomes platform independent and also de-links IMS software implementation modules from data processing scheduler software. XSU is supported in Oracle which would enable retrieval and storage of the data in to a database as an XML file. Report formatting and information content are subjected to change as per requirements of work centers resulting in software version change. Usage of XSL and DTD would result in easier formatting and validation of data and new requirements would not effect software level changes.

The present software modules interact across systems based on RMI calls that are executed synchronously. This results in communication, execution coupling between the sender and the receiver. Event based architectures decouple sender and receiver message exchanges using events which are asynchronous rather than message calls. Implementation of event based architectures requires a messaging server and maintenance of a message queue. Each sender and receiver would have to register for events to be received or raised. An event router would enable messages to be transmitted across geographical boundaries.  Messages from other centers can be received and logged in a centralized server for further transmission or analysis.

The ever growing needs of management to improve data product sales requires knowledge of past data products being sold or generated. To undertake such analysis historic production data has to be archived. Data ware house technology provides a platform for storing historic in lightly summarized form. Building a data warehouse is undertaken in a controlled environment.  A data model is evolved from different perspectives of analysts for storing the data warehouse data. The implementation also requires ETL programs to be developed for populating data warehouse data. Parameters for storing data in data warehouse like wrinkle time to be evolved. A data warehouse technology provides a mechanism for drill down and roll up analysis suitable for extracting knowledge from the data. The mining and extraction of knowledge from data stored in data warehouse would facilitate improved and timely managerial decisions.

\section{Conclusion}
As has been noted earlier, software modules in existing architecture are tightly coupled. Asynchronous message passing mechanism would increase throughput for bulk data products generation as against existing synchronous message passing mechanism. Implementation of XML technologies through databases such as Oracle facilitates better and improved production chain monitoring and realization of interoperable machine independent interfaces with data products generation software. A hybrid architecture encompassing the above technologies streamlines workflow and software management at process level for ever changing business requirements. The benefits of the proposed hybrid architecture are summarized as under:
\begin{enumerate}
\item Service orientation facilitates loose coupling of processes between operating systems, programming languages and other technologies including middleware.
\item	Use of event based message exchange mechanism results in decoupling of the sender and receiver messages.   Event based architectures also facilitates transmission and processing of data across geographical boundaries.  This architecture also facilitates state of art event routing information security mechanisms through subscriber - publisher models.
\item XML based data exchange facilitates platform independence and automated checks on well formed ness of a document and styling for report generation.
\item Support for plug and play processes ensures designer would focus on interfacing and integrating different processes for each activity, thereby eliminating or reducing development time to support new missions.
\item A realization of interoperable machine independent interfaces facilitates design and development and realization of reengineered processes.
\item Replacing existing legacy processes with state of art reengineered processes in a plug and play environment facilitates and streamlines software maintenance activities.
\item	Usage of XSU technology would enable data to stored and retrieved in XML. The report generation software module maintenance is reduced using XSL.
\item Implementations of data warehouse facilitates roll up and drill down analysis of production data that in turn enables business process reengineering leading to improved product generation and customer satisfaction levels by.
\item	The present study facilitates further study in related areas like service discovery, knowledge discovering and event routing.
\end{enumerate}

\%appendices
%
%
%
%
%

\ifCLASSOPTIONcaptionsoff
  \newpage
\fi

\end{document}